\documentclass[12pt]{article}
\usepackage{epsf}
\title{Quark distributions in polarized $\rho$ meson and its 
comparison with those in pion}
\author{A.G. Oganesian\thanks{e-mail address: armen@vitep5.itep.ru}\\
Institute of Theoretical and Experimental Physics,\\
B.Cheremushkinskaya 25, Moscow,117218,Russia}
\date{}
\begin{document}
\maketitle

\newcommand{\be}{\begin{equation}}
\newcommand{\ee}{\end{equation}}

\def\la{\mathrel{\mathpalette\fun <}}
\def\ga{\mathrel{\mathpalette\fun >}}
\def\fun#1#2{\lower3.6pt\vbox{\baselineskip0pt\lineskip.9pt
\ialign{$\mathsurround=0pt#1\hfil##\hfil$\crcr#2\crcr\sim\crcr}}}

\begin{abstract}
Valence quark distributions in transversally and longitudinally 
polarizied $\rho$ mesons in the region of intermediate 
$x$ are obtained by generalizied QCD sum rules. Power corrections 
up to $d$=6 are taken into account. Comparison of the results 
for $\pi$ and $\rho$ mesons shows, that polarization effects are very 
significant and SU(6) symmetry of distribution functions is absent. 
The strong suppression of quark and gluon sea distributions in 
longitudinally polarizied $\rho$ mesons is found.

\end{abstract}

\newpage

\section{Introduction}

Structure functions are one of the most significant  characteristics 
of the inner structure of hadrons . For nucleon and pion there are 
experimental results (see, e.g. \cite{1}-\cite{4} for the 
nucleon, and \cite{5}-\cite{ab}, for the pion). For other hadrons,
one should use some models 
based on additional suppositions about inner structure. A very 
significant question about quark  distribution dependence on 
hadron polarization has no model independent answer 
(moreover in various models it is usually simply supposed that 
there is no significant influence). 

That is why determination of quark  distribution functions in 
a model-independent way in QCD sum rules based only on QCD and the 
operator product expansion (OPE) seems to be very important 
task especially for polarized hadron (in this talk we will discuss 
polarized $\rho$ meson case). 

QCD sum rules for valence quark distribution 
at intermediate $x$ 
was suggested in \cite{7} and developed in \cite{8}-\cite{10}. The method 
was based on the fact, that the imaginary part (in $s$- channel) of 
4-point correlator, corresponding to the forward scattering of a 
virtual photon on the current with the quantum numbers of hadron 
of interest is dominated by contribution of small distances
(at intermediate $x$) if  the virtualities 
of the photon and hadronic 
current ($q^2$ and $p^2$, respectively) are large and 
negative $\vert q^2 \vert
\gg \vert p^2 \vert \gg R^{-2}_c$ , where $R_c$ is the confinement radius.
So the operator product
expansion in this $x$ region is applicable. Then, comparing dispersion 
representation of the forward amplitude in terms of physical states 
with that, calculated in OPE and using Borel transformation 
to suppress higher resonance contributions, one can find quark distribution 
functions at intermediate $x$. 
Unfortunately, the accuracy of the sum rules for nucleon, 
obtained in \cite{8,10}, are bad, (especially for $d$-quark), 
moreover it was found to be impossible to calculate quark distributions in
the $\pi$ and $\rho$ mesons in this way. 
The reason is that the sum rules in
the form used in \cite{8,10} have a serious drawback. 

This drawback comes from the fact that contribution of nondiagonal 
transitions 
wasn't suppressed in sum rules by borelization, and special 
additional procedure, which was used in \cite{8,10} to supress them is 
incorrect for such sum rules. There is no time to discuss this in details, 
the full analisys can be found in our papers \cite{11} and \cite{mn}. 
In this papers the modified method of calculating of the
hadron structure functions (quark distributions in hadrons) 
was suggested. The problem of supressing the nondiagonal terms 
is eliminated and valence quark distributions in pion and 
polarizied $\rho$ meson were calculated.

\section{The method}

Let me briefly present the method.
Consider the non-forward 4-point correlator

$$
\Pi (p_1, p_2; q, q^{\prime}) = -i \int~ d^4x d^4y d^4z e^{ip_1x + iqy -
ip_2z}
$$
\be
\times \langle 0 \vert T \left \{j^h(x),~ j^{el}(y),~ j^{el}(0),~ j^h(z)
\right \} \vert 0 \rangle
\label{3}
\ee
Here $p_1$ and $p_2$ are the initial and final momenta carried by hadronic the 
current $j^h$, $q$ and $q^{\prime} = q + p_1 - p_2$ are the initial and
final momenta carried by virtual photons and Lorentz indices are omitted. It
will be very essential for us to consider unequal $p_1$ and $p_2$ and treat
$p^2_1$ and $p^2_2$ as two independent variables. However, we may put $q^2 =
q^{\prime 2}$ and $t = (p_1 - p_2)^2 = 0$. The general form of the
double dispersion relation (in $p^2_1, p^2_2$) of the imaginary part 
of $\Pi(p^2_1, p^2_2, q^2, s)$ in the $s$ channel has the form

$$
Im \Pi(p^2_1, p^2_2, q^2, s) = a(q^2, s) + \int \limits^{\infty}_0
~\frac{\varphi(q^2, s, u)}{u - p^2_1} du + \int \limits^{\infty} _{0}
\frac {\varphi (q^2, s, u)}{u - p^2_2} du$$
\be
+ \int \limits ^{\infty}_{0} du_1~ \int \limits ^{\infty} _{0} du_2~ \frac
{\rho(q^2, s, u_1, u_2)}{(u_1 - p^2_1) (u_2 - p^2_2)}
\label{4}
\ee
Double Borel
 transformation in $p^2_1$ and $p^2_2$ to (\ref{4}) eliminates
 three first terms and we have

\be
{\cal{B}}_{M^2_1} {\cal{B}}_{M^2_2}~ Im \Pi(p^2_1, p^2_2, q^2, s) = \int
\limits ^{\infty} _{0} du_1~ \int \limits ^{\infty}_{0} du_2 \rho (q^2, s,
u_1, u_2) exp \Biggl [-\frac{u_1}{M^2_1} - \frac{u_2}{M^2_2} \Biggr ]
\label{5}
\ee
where $M^2_1$ and $M^2_2$ are the squared Borel mass.
The integration region with respect to  $u_1, u_2$ may be divided 
into four areas:

I. $u_1 < s_0,~~ u_2 < s_0;$

II.$u_1 < s_0,~~ u_2 > s_0;$

III. $u_1 > s_0, ~~ u_2 < s_0;$

IV. $u_1, u_2 > s_0.$

Here  $s_0$ is the continuum threshold in the standard QCD sum rule 
model of the hadronic spectrum with one lowest resonance plus continuum.
Area I obviously corresponds to the resonance
contribution and the spectral density in this area can be written as

\be
\rho(u_1, u_2, x, Q^2) = g^2_h \cdot 2\pi F_2 (x, Q^2) \delta(u_1 - m^2_h)
\delta(u_2 - m^2_h),
\label{8}
\ee
where $g_h$ is defined as

\be
\langle 0 \vert j_h \vert h \rangle = g_h
\label{9}
\ee
(For simplicity we consider the case of the Lorentz scalar hadronic current.
The necessary modifications  for the $\pi$ and $\rho$ mesons will be
presented below). If the structure,
proportional to $P_{\mu} P_{\nu}$ $[P_{\mu} = (p_1 + p_2)_{\mu}/2]$ is
considered in $Im \Pi(p_1, p_2, q, q^{\prime})$, then in the lowest 
twist approximation $F_2(x, Q^2)$ is the structure function of interest. 

In area IV, where both variables $u_{1,2}$ are far from the resonance region,
the nonperturbative effects may be neglected and, as usual in the sum rules,
the spectral function of a hadron state is described by the bare loop spectral
function $\rho^0$ in the same region:

\be
\rho(u_1, u_2, x) = \rho^0(u_1, u_2, x)
\label{8}
\ee
In areas II and III one of the variables is far from the resonance region,
but another is in the resonance region, and the spectral function in this
region is some unknown function $\rho = \psi(u_1, u_2, x)$, which
corresponds to the transitions like $h \to continuum$. 

Taking all these facts into account, the physical side of the sum 
rule (\ref{5}) can be rewritten as 

$$
\hat{B}_1 \hat{B}_2 [Im \Pi] = 2 \pi F_2 (x, Q^2) \cdot g^2_h e^{-m^2_h
(\frac{1}{M^2_1}+ \frac{1}{M^2_2})} + \int \limits^{s_0} _{0} du_1 ~\int
\limits ^{\infty}_{s_0}~ du_2 \psi (u_1, u_2, x) e^{-(\frac{u_1}{M^2_1} +
\frac{u_2}{M^2_2})}
$$
\be
+\int \limits^{\infty} _{s_0} du_1 ~\int
\limits ^{s_0}_0~ du_2 \psi (u_1, u_2, x) e^{-(\frac{u_1}{M^2_1} +
\frac{u_2}{M^2_2})} + \int \limits ^{\infty}_{s_0} \int \limits ^{\infty}
_{s_0} du_1 du_2 \rho^0 (u_1, u_2, x)e^{-(\frac{u_1}{M^2_1} +
\frac{u_2}{M^2_2})}
\label{9}
\ee
In what follows, we put $M^2_1 = M^2_2 \equiv 2 M^2$.
(As was shown in \cite{12}, the values of the Borel parameters $M^2_1,
M^2_2$ in the double Borel transformation are about twice as large as those 
in the ordinary ones).

One of the advantages of this method is that after the 
double Borel transformation the unknown contributions of II and III areas
[the second and the third terms in (\ref{9})] are exponentially suppressed. 
So, and we would like to emphasise this, we do not need in any additional 
artificial procedure such as the differentiation with respect to the  
Borel mass.
Using standard duality arguments, we estimate the contribution 
of all nonresonance region (i.e., areas II, III, IV) as the  
contribution of the bare loop in the
same region and demand their value to be small (less than 30\%). Finally,
equating the physical and QCD representations of $\Pi$ one can
write the following sum rules:

$$
Im~ \Pi^0_{QCD} + \mbox{Power~ correction} = 2 \pi F_2 (x, Q^2) g^2_h e
^{-m^2_h(\frac{1}{M^2_1} + \frac{1}{M^2_2})}
$$
\be
Im \Pi^0_{QCD} = \int \limits^{s_0}_{0}\int \limits^{s_0}_{0}\rho^0 (u_1,
u_2, x) e^{- \frac{u_1 + u_2}{2M^2}}
\label{10}
\ee

It is worth mentioning that if we consider the forward
scattering amplitude from the very beginning, (i.e. put $p_1 = p_2 = p$ 
as in \cite {7}-\cite {10}) and perform the 
Borel transformation in $p^2$, then the contributions of
the second and third terms in (\ref{4}),in contrast to (\ref{9}), 
would not be suppressed comparing
with the lowest resonance contribution we are interesting in. They 
correspond to the nondiagonal transition matrix elements discussed in the
Introduction.


\section{ Quark distributions in pion}

Let me briefly discuss the main points of the calculation and the 
results for the pion structure function, 
which can be treated as a check of the accuracy of the method due to 
fact that for pion the experimental results are avaible.
To find the pion structure function by the method, described in the 
previous section, one should consider the imaginary part of 4-point 
correlator (\ref{3}) with two axial and two electromagnetic currents.
Since $\bar{d}(x) = u(x)$, it is enough to find the distribution 
of the valence $u$ quark in $\pi^+$. 
The most suitable choice of the axial current is

\be
j_{\mu 5} = \bar{u} \gamma_{\mu} \gamma_5 d
\label{11}
\ee
The electromagnetic current is chosen as $u$-quark current 
with the unit charge

\be
j^{el}_{\mu} = \bar{u} \gamma_{\mu} u
\label{12}
\ee

The most convenient tensor structure, which is chosen to construct the 
sum rule, is a 
structure proportional to $P_{\mu} P_{\nu} P_{\lambda} P_{\sigma}/\nu$. 

The sum rule for the valence $u$-quark distribution in the pion in 
the bare loop approximation is \cite{11}:

\be
u_{\pi}(x) = \frac{3}{2 \pi^2}~\frac{M^2}{f^2_{\pi}} x (1-x) (1 -
e^{-s_0/M^2}) e^{m^2_{\pi}/M^2},
\label{16}
\ee
where $s_0$ is the continuum threshold.
In \cite{11} the following corrections to Eq. (\ref{16}) were 
taken into account:

1. Leading order (LO) perturbative corrections proportional to
$ln(Q^2/\mu^2)$ , where $\mu^2$ is the normalization point. In what follows,
the normalization point will be chosen to be equal to the Borel parameter
$\mu^2 = M^2$.

2.  Power corrections -- higher order terms of OPE. Among them, the
dimension-4 correction, proportional to the gluon condensate $\langle 0 \vert
\frac{\alpha_s}{\pi} G^n_{\mu \nu}~ G^n_{\mu\nu} \vert 0 \rangle$ was first
taken into account, but it was found that the gluon condensate contribution 
to the
sum rule vanishes after double borelization. There are two types of vacuum
expectation values of dimension 6. One, involving only gluonic fields:

\be
\frac{g_s}{\pi} \alpha_s f^{abc} \langle 0 \vert G^{a}_{\mu \nu}~ G^b_{\nu
\lambda}~ G^c_{\lambda \mu} \vert 0 \rangle
\label{17}
\ee
and the other, proportional to the four-quark operators

\be
\langle 0 \vert \bar{\psi} \Gamma \psi \cdot \bar{\psi} \Gamma \psi \vert 0
\rangle
\label{18}
\ee
It was shown in \cite{11} that terms of the first type cancel in the sum
rule and only terms of the second type survive. For the latter, one may use
the factorization hypothesis which reduces all the terms of this type to the
square of the quark condensate.

 A remark is in order here. As was mentioneed in the Introduction, the 
present approach is invalid at small and large $x$. No-loop 4-quark 
condensate contributions are proportional to $\delta(1-x)$ and, 
being outside of the applicability domain of the approach, cannot be 
taken into account. In the same way, the diagrams, which can be 
considered as 
radiative corrections to those, proportional to $\delta(1-x)$ must be also
omitted.

All dimension-6 power corrections to the sum rule were calculated in 
\cite{11} and the final result is given by (the pion mass is neglected):

$$
xu_{\pi}(x) = \frac{3}{2\pi^2}\frac{M^2}{f^2_{\pi}}x^2(1-x)
\Biggl [ \Biggl ( 1+ \Biggl (\frac{a_s(M^2)\cdot ln(Q^2_0/M^2)}{3\pi}\Biggr
)$$

$$\times \Biggl ( \frac{1+4x ln(1-x)}{x}- \frac{2(1-2x)ln x}{1-x}\Biggl )
\Biggl )\cdot (1-e^{-s_0/M^2}) $$

\be
-\frac{4\pi \alpha_s(M^2)\cdot 4\pi \alpha_s a^2}{(2\pi)^4 \cdot
3^7\cdot 2^6\cdot M^6} \cdot \frac{\omega(x)}{x^3(1-x)^3}\Biggr ],
\label{19}
\ee
where $\omega(x)$ is the fourth degree polynomial in $x$,

\be a =
-(2 \pi)^2 \langle 0 \vert  \bar{\psi} \psi \vert 0 \rangle
\label{20}
\ee

$$
\omega(x) = -5784x^4 - 1140x^3 - 20196x^2
$$
$$
+ 20628x - 8292) ln(2) + 4740x^4 + 8847x^3
$$
\be
+ 2066x^2 - 2553x + 1416
\label{21}
\ee
The function $u_{\pi}(x)$ may be used as an initial condition at $Q^2 = Q^2_0$
for solution of
QCD evolution equations (Dokshitzer-Gribov-Lipatov-Altarelli-Parisi 
equations). 

In the numerical calculations we choose: the effective  
$\Lambda^{LO}_{QCD} = 200 ~MeV,~ Q^2_0 = 2 ~GeV^2, ~ \alpha_s a^2 (1 GeV^2) =
0.13 ~GeV^6$ \cite{11}. The continuum threshold was varied in the interval
$0.8 < s_0 < 1.2 ~GeV^2$ and it was found, that the results depend only 
slightly 
on it's variation. The analysis of the sum rule (\ref{19}) shows, that it is
fulfilled in the region $0.15 < x < 0.7$; the power corrections are less
than 30\% and the continuum contribution is small ($< 25$\%). The stability
in the Borel mass parameter $M^2$ dependence in the region $0.4 GeV^2 < M^2
< 0.6 GeV^2$ is good. The result of our calculation of the valence quark 
distribution $x u_{\pi}(x,Q^2_0)$ in the pion is shown  in Fig. 1.
In Fig. 1, we
plot also the valence $u$-quark distribution found in \cite{6} by
fitting the data on the production of $\mu^+\mu^-$ and $e^+e^-$ pairs in
pion-nucleon collisions (Drell-Yan process). When comparing with the 
distribution found here it should be remembered, that the accuracy of our curve is of
order of $ 10 - 20\%$, (see discussion in Section 5). The $u$-quark 
distribution 
found from the experiment is also not free from uncertainties (at least 10-20\%, 
see \cite{5} - \cite {ab}). 
For all these reasons, we consider the agreement of
two curves as being good. 

Assume, that $u_{\pi}(x) \sim 1/\sqrt{x}$ at small $x \la 0.15$  
according
to the Regge behaviour and $u_{\pi}(x) \sim (1-x)^2$
at large $x \ga 0.7$
according to quark counting rules. Then, matching these functions with
(\ref{19}), one may find the numerical values of the first and the second
moments of the $u$-quark distribution:

\be
{\cal{M}}_1 = \int \limits^1 _0 u_{\pi} (x) dx \approx 0.84 ~~ (0.85)
\label{22}
\ee

\be
{\cal{M}}_2 = \int \limits^1_0 xu_{\pi} (x) dx \approx 0.21 ~~ (0.23)
\label{23}
\ee
where the values in the parentheses correspond to behavior 
$u_{\pi}(x) \sim (1- x)$ at large $x$. The results depend only slightly 
on the matching points (not more than 5\%, when the lower matching 
point is varied in the region 0.15 - 0.2 and the upper one in the 
region 0.65 - 0.75). The moment ${\cal{M}}_1$ has the meaning of 
the number of $u$ quarks in $\pi^+$ and it
should be ${\cal{M}}_1 = 1$. The deviation of (\ref{22}) from 1 characterizes
the accuracy of our calculation. The moment ${\cal{M}}_2$ has the meaning 
of the pion momentum fraction carried by the valence $u$ quark. Therefore, 
the valence $u$ and
$\bar{d}$ quarks carry about 40\% of the total momentum. 

The valence $u$-quark distribution in the pion 
was calculated recently in the instanton model \cite{13}. At intermediate $x$,
the values of $xu_{\pi}(x)$ found in \cite{13} are not more than 20\%
higher that our results.
Recently, the pions valence quark momentum distribution using a model, 
based on the Dyson-Schwinger equation was also calculated \cite{cd}.
Our results are in a reasonable agreement
with the results of  \cite{cd}. Our estimation (\ref{23}) of the second 
moment of the valence quark distribution 
can be also compared 
with the calculation of the second moment of the total  (valence plus sea) 
quark distribution  in the pion  \cite{ef}. The value, 
obtained in \cite {ef}, is 0.6 for the total momentum carried by all quarks 
(valence plus sea) 
with the accuracy about 10\%. Taking into account, that sea quarks usually 
supposed to carry 15\% of the total momentum, one can estimate from the result
of \cite{ef}, that the second moment of one valence quark distribution 
should be about 
20-22\%, which is in a good agreement with our result (\ref{23}). 
The quark distribution in the pion was calculated also in \cite{kl} by using sum 
rules with nonlocal condensates. Unfortunately it is impossible to 
perform the 
comparison directly, because the quark distribution is calculated in \cite{kl} 
only at very low  normalization point. But, comparig his result with 
different models, the author of this paper
arrived at the conclusion, that the result is in agreement with 
experimental data at $Q^2$=20 $GeV^2$ within the accuracy about 20\%. 
Our result is close also to experimental fit, so we can believe 
that our results are in an agreement with those of \cite{kl}.

\section{Quark distributions in $\rho$ meson}

Let us calculate valence $u$-quark distribution in the $\rho^+$ meson. 
The choice of hadronic current is evident

\be
j_{\mu}^{\rho} = \overline{u}\gamma_{\mu}d
\label{24}
\ee
The matrix element $\langle \rho^+\mid j^{\rho}_{\mu}\mid 0 \rangle$ is
given by

\be
\langle \rho^+\mid j^{\rho}_{\mu}\mid 0 \rangle =
\frac{m^2_{\rho}}{g_{\rho}}e_{\mu}
\label {25}
\ee
where $m_{\rho}$ is the $\rho$-meson mass, $g_{\rho}$ is the $\rho-\gamma$
coupling constant, $g^2_{\rho}/4\pi=1.27$ and $e_{\mu}$  is the $\rho$ meson
polarization vector. Consider the coordinate system, where the collision 
of the 
$\rho$-meson with momentum $p$ and the virtual photon with momentum $q$
proceeds along $z$-axes. Averaging over $\rho$ polarizations is given by
the formulas:\\

\be
e^L_{\mu}e^L_{\nu}=
\Biggl ( q_{\mu} - \frac{\nu p_{\mu}}{m^2_{\rho}}\Biggr )
\Biggl ( q_{\nu} - \frac{\nu p_{\nu}}{m^2_{\rho}}\Biggr )
\frac{m^2_{\rho}}{\nu^2-q^2m^2_{\rho}}
\label{26}
\ee
for longitudinally polarized $\rho$ and

\be
\sum_{T,r}e^r_{\mu}e^r_{\nu} = -\Biggl ( \delta_{\mu\nu} -
\frac{p_{\mu}p_{\nu}}{m^2_{\rho}} \Biggr ) -
\frac{m^2_{\rho}}{\nu^2-q^2m^2_{\rho}}
\Biggl ( q_{\mu}- \frac{\nu p_{\mu}}{m^2_{\rho}} \Biggr )
\Biggl ( q_{\nu}- \frac{\nu p_{\nu}}{m^2_{\rho}} \Biggr )
\label{27}
\ee
for transversally polarized $\rho$.

The imaginary part of the forward $\rho-\gamma$  scattering amplitude
$W_{\mu\nu\lambda\sigma}$
(before multiplication by $\rho$ polarizations)  satisfies the equations
$W_{\mu\nu\lambda\sigma} q_{\mu}=$ $W_{\mu\nu\lambda\sigma} q_{\nu}=$
$W_{\mu\nu\lambda\sigma} p_{\lambda}=$ $W_{\mu\nu\lambda\sigma}
p_{\sigma}=0$, which follow from current conservation. The indices
$\mu,\nu$ refer to the initial and final photon; $\lambda,\sigma$ -- to 
the initial
and final $\rho$.  The general form of $W_{\mu\nu\lambda\sigma} $  is:

$$W_{\mu\nu\lambda\sigma} = \Biggl [\Biggl
( \delta_{\mu\nu} - \frac{q_{\mu}q_{\nu}}{q^2}\Biggr ) \Biggl (
\delta_{\lambda\sigma}-\frac{p_{\lambda}p_{\sigma}}{m^2_{\rho}}\Biggr ) A-
\Biggl ( \delta_{\mu\nu} - \frac{q_{\mu}q_{\nu}}{q^2}\Biggr ) \Biggl (
q_{\lambda}- \frac{\nu p_{\lambda}}{m^2_{\rho}}\Biggr ) \Biggl (
q_{\sigma}-\frac{\nu p_{\sigma}}{m^2_{\rho}}\Biggr ) B$$

$$ - \Biggl ( p_{\mu} - \frac{\nu q_{\mu}}{q^2}\Biggr )
\Biggl ( p_{\nu} - \frac{\nu q_{\nu}}{q^2}\Biggr ) \Biggl (
\delta_{\lambda\sigma} -\frac{p_{\lambda} p_{\sigma}}{m^2_{\rho}} \Biggr ) C
+ \Biggl ( p_{\mu} - \frac{\nu q_{\mu}}{q^2}\Biggr )
\Biggl ( p_{\nu} - \frac{\nu q_{\nu}}{q^2}\Biggr )$$

\be
\times \Biggl ( q_{\lambda} - \frac{\nu p_{\lambda}}{m^2_{\rho}}\Biggr )
\Biggl ( q_{\sigma} - \frac{\nu p_{\sigma}}{m^2_{\rho}}\Biggr ) D \Biggr ]
\label{28}
\ee
where A,B,C and D are the invariant functions. By averaging  
Eq. (\ref{28}) over
polarizations for longitudinal and transverse $\rho$ mesons, one can find that 
the structure function $F_2(x)$ proportional to $p_{\mu}p_{\nu}$ 
is given in the
scaling limit $(\nu^2 \gg \mid q^2 \mid m^2_{\rho})$ by the
contribution of invariants $C+(\nu^2/m^2_{\rho})D$  and $C$ 
in the cases of the longitudinal
and the transversal $\rho$ meson, respectively. This means that, 
in the forward scattering amplitude $W_{\mu\nu\lambda\sigma}$ (\ref{28}), one
must separate the structure proportional to
$p_{\mu}p_{\nu}p_{\lambda}p_{\sigma}$ in the first case and the structure
$\sim p_{\mu}p_{\nu}\delta_{\lambda\sigma}$ in the second case.

Consider now the non-forward 4-point correlator

$$\Pi_{\mu\nu\lambda\sigma} (p_1,p_2;q,q^{\prime}) = -i\int d^4 x d^4 y d^4
ze^{ip_1x+iqy-ip_2z} $$

\be
\times \langle 0\mid T~\{ j^{\rho}_{\lambda}
(x),~j^{el}_{\mu}(y),~j^{el}_{\nu}(0),~j^{\rho}_{\sigma}(z)\}\mid 0
\rangle,
\label{29}
\ee
where the currents $j^{el}_{\mu}(x)$ and $j^{\rho}_{\lambda}(x)$ are given
by Eq.(\ref{12}) and (\ref{25}). It is evident from the consideration, 
that in the non-forward amplitude 
the most suitable tensor structure for determination of 
$u$-quark distribution in the longitudinal $\rho$ meson 
is that
proportional to $P_{\mu} P_{\nu} P_{\sigma}P_{\lambda}$, while $u$-quark
distribution in the transverse $\rho$  can be found by considering 
the invariant
function at the structure ($-P_{\mu}P_{\nu}\delta_{\lambda\sigma}$). 

In the case of longitudinal $\rho$ meson the tensor structure, that 
is separated  is the
same as in the case of the pion. Since bare loop contributions for
vector and axial hadronic currents are equal at $m_q=0$, the only 
difference from the
pion case is the normalization. It can be shown, that $u$-quark
distribution in the longitudinal $\rho$ meson can be found from 
Eq.(\ref{19}) by
substituting $m_{\pi}\to m_{\rho}$, $f_{\pi}\to m_{\rho}/g_{\rho}$ and,
therefore, one can easily write down sum rules for this distribution:

$$xu^L_{\rho}(x) = \frac{3}{2\pi^2}
M^2\frac{g^2_{\rho}}{m^2_{\rho}}e^{m^2_{\rho}/M^2} x^2(1-x)\Biggl [\Biggl
(1+ \Biggl (\frac{a_s(M^2)\cdot ln(Q^2_0/M^2)}{3\pi}\Biggl )$$

\be
\times \Biggl (\frac{1+4x ln(1-x)}{x} - \frac{2(1-2x)ln x}{1-x}\Biggr )
\Biggr )(1-e^{-s_0/M^2})
-\frac{\alpha_s(M^2)\cdot \alpha_s a^2}{\pi^2 \cdot 3^7\cdot 2^6\cdot
M^6}\cdot \frac{\omega(x)}{x^3(1-x)^3}\Biggr ],
\label{30}
\ee
where $a$  and $\omega(x)$  are given by Eqs.(\ref{20}) and (\ref{21}), 
respectively. Sum rules for $u^L_{\rho}(x)$  are
satisfied in the wide $x$ region:  $0.1 < x < 0.85$. The Borel mass $M^2$
dependence is weak in the whole range of $x$, except $x\leq 0.15$ and  $x
\geq 0.7$. 
Figure. 2 presents $xu^L_{\rho}(x)$  as a function of
$x$. The values $M^2=1$ GeV$^2$ and $s_0=1.5$ GeV$^2$, $Q_0^2=4$ GeV$^2$  were
chosen, the parameters $\Lambda^{LO}_{QCD}$ and $\alpha_s
a^2$  are the same as in the calculation of $xu_{\pi}(x)$.

Let us now consider the case of transversally polarizied  $\rho$-meson, 
i.e., the  term proportional to the
structure $P_{\mu}P_{\nu}\delta_{\lambda\sigma}$. The 
bare loop  contribution (with leading order perturbative corrections) 
is  

$$u^T (x) = \frac{3}{8\pi^2}
\frac{g^2_{\rho}}{m^4_{\rho}}e^{m^2_{\rho}/M^2}\cdot M^4 \cdot E_1\Biggl
(\frac{s_0}{M^2}\Biggr ) \cdot \varphi_0(x)$$

\be
\Biggl [ 1 + \frac{ln(Q^2_0/\mu^2)\cdot \alpha_s(\mu^2)}{3\pi}\cdot
\Biggl (
(4x-1)/\varphi_0(x) + 4ln(1-x) - \frac{2(1-2x+4x^2)lnx}{\varphi_0(x)}\Biggr )
\Biggr ]
\label{34}
\ee
where

\be
\varphi_0(x)=1-2x(1-x)
\label{35}
\ee

We consider now the power correction contribution to the sum rules. 
The power correction of the lowest dimension is
proportional to the gluon condensate
$\langle G^q_{\mu\nu}G^q_{\mu\nu}\rangle$ with $d=4$. 
The
$\langle G^q_{\mu\nu}G^q_{\mu\nu}\rangle$ correction was calculated in the
standard way  in the Fock-Schwinger gauge
$x_{\mu}A_{\mu}=0$ \cite{15}.

The quark propagator $iS(x,y)=\langle \psi(x)\psi(y)\rangle$ in the external
field $A_{\mu}$ has the well-known form \cite{12,8,15,16}. 
In contrast to the pion case, the $\langle G^a_{\mu\nu}G^a_{\mu\nu}\rangle$ 
correction for transversally polarized $\rho(\rho_T)$, 
does not vanish. 

\be
Im\Pi^{(d=4)}_T = -\frac{\pi}{8x}\langle 0\mid\frac{\alpha_s}{\pi}
G^2_{\mu\nu}\mid 0\rangle
\label{36}
\ee

There are a great number of loop diagrams for $d=6$ correction. It is
convenient to divide  them into two types and discuss these types
separately. Type I diagrams are those, in which only the interaction 
with the external gluon field is taken into account and
type II diagrams are those in which expansion of the quark field 
is also taken into account.

Discuss briefly the
special features of calculating diagrams of these two types. The 
type-I diagrams are obviously proportional  to $\langle 0 \mid g^3
f^{abc}G^a_{\mu\nu}G^b_{\alpha\beta}G^c_{\rho\sigma} \mid 0 \rangle $, 
$\langle 0 \mid
D_{\rho}G^a_{\mu\nu}D_{\tau}G^a_{\alpha\beta}\mid 0\rangle$
and  $\langle 0 \mid
G^a_{\mu\nu}D_{\rho}D_{\tau}G^a_{\alpha\beta}\mid 0\rangle$.

One may demonstrate \cite{17} that these tensor structures are
proportional to two vacuum averages

$$\langle 0 \mid g^2j^2_{\mu} \mid 0 \rangle ~~~\mbox{and}~~~
\langle 0 \mid g^3 G^a_{\mu\nu} G^b_{\nu\rho} G^c_{\rho\mu}f^{abc} \mid 0
\rangle.$$
By using the factorization hypothesis 
the first of them, $\langle 0 \mid g^2j^2_\mu\mid 0\rangle$ 
easily reduces to $\langle g\bar{\psi}\psi\rangle^2$, which is well known:

\be
\langle 0 \mid g^2j^2_{\mu}\mid 0 \rangle = -(4/3)[\langle 0 \mid
g\bar{\psi}\psi \mid 0 \rangle ]^2.
\label{39}
\ee
But $\langle 0 \mid g^3
G^a_{\mu\nu} G^b_{\nu\rho} G^c_{\rho\mu}f^{abc} \mid 0 \rangle$ is not well
known; there are only some estimates based on the instanton model
\cite{18,19}. In contrast to $\pi$ and $\rho_L$-meson cases, the terms 
proportional to\\
$\langle 0\mid  g^3 f^{abc} G^a_{\mu\nu}G^b_{\nu\rho}$ $G^c_{\rho\mu} \mid 0
\rangle$  are not cancelled for $\rho_T$ and one should
estimate it. The
estimation based on the instanton model \cite{18} gives

\be
-\langle g^3 f^{abc}G^a_{\mu\nu} G^b_{\nu\rho} G^c_{\rho\mu}\rangle =
\frac{48\pi^2}{5} \frac{1}{\rho^2_c} \langle 0\mid(\alpha_s/\pi) G^2_{\mu\nu}
\mid 0\rangle,
\label{40}
\ee
where $\rho_c$  is the effective instanton radius.

Among type-II diagrams only those, in which the
interaction with the vacuum takes place inside the loop are considered. Such
diagrams cannot be treated as the evolution of any non-loop diagrams and are
pure power corrections of dimension 6. 

The total number of $d=6$ diagrams is enormous -- about 500. Collecting the
results we get finally the following sum rules for the valence 
$u$-quark distribution in the transversally polarized $\rho$ meson.

$$
xu^T_{\rho}(x) = \frac{3}{8 \pi^2} g^2_{\rho} e^{m^2_{\rho}/M^2}~
\frac{M^4}{m^4_{\rho}} x \left \{\varphi_0(x) E_1 \Biggl (\frac{s_0}{M^2}
\Biggr ) \Biggl [ 1 +
\frac{1}{3 \pi} ln \Biggl (\frac{Q^2_0}{M^2} \Biggr ) \alpha_s (M^2) \Biggl
( \frac{(4x-1)}{\varphi_0(x)} + \right.  $$

$$ + 4 ln (1-x) - \frac{2(1 - 2x
+ 4x^2) ln x}{\varphi_0(x)} \Biggr ) \Biggr ]- \frac{\pi^2}{6}~
\frac{\langle 0 \vert(\alpha_s/\pi) G^2 \vert 0 \rangle} {M^4 x^2}  $$

$$
+\frac{1}{2^8 \cdot 3^5 M^6 x^3 (1-x)^3} \langle 0 \vert
g^3 f^{abc} G^a_{\mu \nu} G^b_{\nu \lambda} G^c_{\lambda \mu} \vert 0
\rangle \xi (x)
$$

\be
\left. +\frac{\alpha_s(M^2) (\alpha_s a^2)}{2^5 \cdot 3^8 \pi^2 M^6 x^3 (1
- x)^3 }\chi (x) \right\}
\label{42}
\ee
where 
 \be
 \xi(x) = -1639 + 8039 x - 15233 x^2 +
10055 x^3 - 624x^4 - 974 x^5
\label{43}
\ee

$$
\chi(x) = 8513 - 41692 x + 64589 x^2 - 60154 x^3 + 99948 x^4
$$
$$
- 112516 x^5 + 45792 x^6 + (- 180 - 8604 x + 53532 x^2
$$
\be
- 75492 x^3 - 28872x^4 + 109296 x^5 - 55440 x^6) ln 2
\label{44}
\ee
The standard value of the gluonic  condensate $\langle 0 \mid
(\alpha_s/\pi)G^2\mid 0 \rangle=0.012~GeV^4$ was taken in numerical
calculations. Equation (\ref{40}) was used and 
the effective instanton radius $\rho_c$ was chosen as
$\rho_c = 0.5 fm$. This value is between the estimations of \cite{18}
($\rho_c = 1 fm$) and \cite{19} ($\rho_c = 0.33 fm$). (In the recent paper
\cite{21} it was argued that the liquid gas instanton model
overestimates higher order gluonic condensates and, in order to correct this
effect, larger values of $\rho_c$ comparing with \cite{19} should be used).
The Borel mass dependence of $xu_T(x)$ in the interval $0.2 < x < 0.65$ 
is weak at $0.8 < M^2 < 1.2 ~GeV^2$. Figure 3 shows $xu_T(x)$ at $M^2 = 1
~GeV^2$ and $Q^2_0 = 4 ~GeV^2$.  Dashed and thin solid lines demonstrate 
the influence of
the variation in $\rho_c$ on the final result:  the lower line corresponds
to $\rho_c = 0.6 fm$ and the upper -- to $\rho_c = 0.4 fm$.  Our results are
reliable at $0.2 < x < 0.65$, where $d = 4$ and $d = 6$ power
corrections each comprise less than 30\% of the bare loop contribution. 
(The contributions $\langle
0 \vert (\alpha_s/\pi) G^2 \vert 0 \rangle$ and $\langle 0 \vert g^3 f^{abc}
G^a_{\mu\nu}G^b_{\nu\lambda}G^c_{\lambda\mu} \vert 0 \rangle$ 
are of the opposite sign and compensate one another, $\alpha_s(M^2)\alpha_s
a^2$ contribution is negligible.) At $\rho_c = 0.4 fm$ the applicability
domain shrinks to $0.25 < x < 0.6$. 

The moments of quark distributions in the 
longitudinal $\rho$ meson are calculated in the same way, as it 
was done in
for the case of pion: by matching with Regge behavior 
$u(x) \sim 1/\sqrt{x}$ at
low $x$ and with quark counting rule $u(x) \sim (1 - x)^2$ at large $x$. The
matching points were chosen as $x = 0.10$ at low $x$ and $x = 0.80$ at large
$x$. The numerical values of the moments for the longitudinally polarized 
$\rho$ are

$$
{\cal{M}}^L_1 = \int \limits ^{1}_{0} dx u^L_{\rho} (x) = 1.06 ~~~ (1.05)
$$
\be
{\cal{M}}^L_2= \int \limits ^{1}_{0} x dx u^L_{\rho} (x) = 0.39 ~~~ (0.37)
\label{45}
\ee
The values of moments, obtained by assuming that $u(x) \sim (1 - x)$ at
large $x$ are given in the parentheses.

Reliable calculation of moments for the case of transversally polarized
$\rho$ meson is impossible, because of a narrow applicability domain in $x$
and expected double humps shape of the $u$-quark distribution, 
which does not allow soft
matching with expected behaviour $xu^T_{\rho}(x)$ at small and large $x$.

Now let us discuss the nonpolarizied $\rho$-meson case. The 
quark distribution function $u(x)$ in this case is equal to

$$u_{\rho}(x)= (u^L_{\rho}(x) + 2 u^T_{\rho}(x))/3$$
and we can determine $u(x)$ only in the region, where sum rules  for
$u^L_{\rho}(x)$ and $u^T_{\rho}$
are satisfied, i.e. in $0.2 \la x\leq 0.65$.
In this region, $u_{\rho}(x)$ is found to be very close to $u_{\pi}(x)$
(the difference in whole range of $x$ is no more than 10-15\%).

\section{Summary and discussion}

\hspace{5mm} Figure 4 gives the comparison of the valence $u$-quark 
distributions in the pion and
longitudinally and transversally polarized $\rho$ mesons. The shapes of
the curves  are quite different, especially of $xu^T_{\rho}(x)$  in comparison
with $xu^L_{\rho}(x)$ and $xu_{\pi}(x)$. 
Strongly different are also the second moments
in the pion and longitudinal $\rho$-meson: the momentum fraction,
carried by valence quarks and antiquarks $-(u+\bar{d})$ in the longitudinal
$\rho$ meson is about 0.8, while in the pion it is much less -- about 0.4-0.5.
All these differences are very large and many times larger than estimated 
uncertainties of our results. In the case of
$u$-quark distribution in the pion the main source of them is the value of the 
renorminvariant quantity $(2\pi)^4
\alpha_s \langle \mid \bar{\psi}\psi \mid 0 \rangle^2$. In our calculations,
we took it to be equal to 0.13 GeV$^6$. In fact, however, it is uncertain by a
factor of 2. (Recent determination \cite{22} of this quantity from
$\tau$-decay data indicates that it may be two times larger). The
perturbative corrections also introduce some uncertainties, especially at large
$x(x > 0.6)$ where the LO correction, which is taken into account, is large. 
The estimation of both effects shows that they may result in 10-20\%
variation (increase at $x < 0.3$ and decrease at $x >0.3$) of
$xu_{\pi}(x)$

For the $u$-quark distribution in the longitudinally polarized $\rho$ meson, 
the uncertainties in $\alpha_s\langle 0\mid \bar{\psi}\psi \mid 0 \rangle^2$
do not play any role, because of higher $M^2$ values, so the expected 
accuracy is even better.
One should note, that the accuracy of prediction of moments for 
longitudally polarizied $\rho$ 
meson are very high, because, as one can see from Fig.4, sum rules predict 
quark distribution almost in whole region of $x$ (from 0.1 up to 0.9). 
The fact that the first moment is so close to 1, also confirm that 
accuracy is high. Large value of the second moment mean, that valence 
quarks in longitudally polarizied $\rho$ meson 
carried about 0.8 of the  
total momentum, so one can conclude that the total (gluon and quark) sea 
is strongly suppressed. Moreover, this prediction for the second moment 
is close to those obtained in \cite{23} from quite another sum rules, 
and also agreed with the lattice calculation results \cite{24}, so the sea 
supression in longitudally polarizied $\rho$ meson can be treated as a
theoretically well established fact.

The accuracy of our results for the $u$-quark distribution in the 
transversally
polarized $\rho$ meson is lower, because of a large role of $d=4$  and $d=6$
gluonic condensate contributions. For the latter, as was discussed before, 
there are only estimations, based on the instanton model, and 
the
gluon condensate $\langle 0 \mid(\alpha_s/\pi)G^2 \mid 0 \rangle$  is also
uncertain by a factor 1.5. Estimations show that they result in not more 
than 30-40\% variation in
$xu_{\rho}^T$  at $x\approx 0.3-0.4$, but in much less variation at $x\approx
0.5-0.6$. (The LO perturbative corrections are
not more than 20\% at small $x$ and negligible at large $x$). 
So one can see, that the difference, obtained in the quark distribution in 
the pion, 
longitudinally and transversally polarizied $\rho$-meson 
in fact are much more larger than any possible 
uncertainties of the results. 

In summary the main physical conclusion are the following: 

(1) The sea is strongly supressed in the longitudinally polarizied 
$\rho$-meson.

(2) The quark distribution in the polarizied $\rho$-meson are significantly  
dependend on polarization. 

(3) The quark distributions in the the pion and
$\rho$-meson have a not too much in common. The specific properties of the pion, as
a Goldstone boson, manifest themselves in
the quark distributions different from those
in the $\rho$ meson. $SU(6)$  symmetry may, probably, take place for
the static properties of $\pi$ and $\rho$, but not for their inner 
structure.
We have no explanation, why $u$-quark
distributions in the pion and unpolarized $\rho$-meson at $0.2 < x < 0.65$ are
close to one another -- whether it is a pure accident or there are 
some deep reasons for it.

This study was supported in part 
by Award no. RP2-2247 of U.S.  Civilian Research and Development Foundation
for Independent  States of Former Soviet Union (CRDF), by the Russian 
Foundation of Basic Research, project no. 00-02-17808 and by INTAS Call
2000, project 587.

\newpage

\begin{figure}
\epsfxsize=5cm
\epsfbox{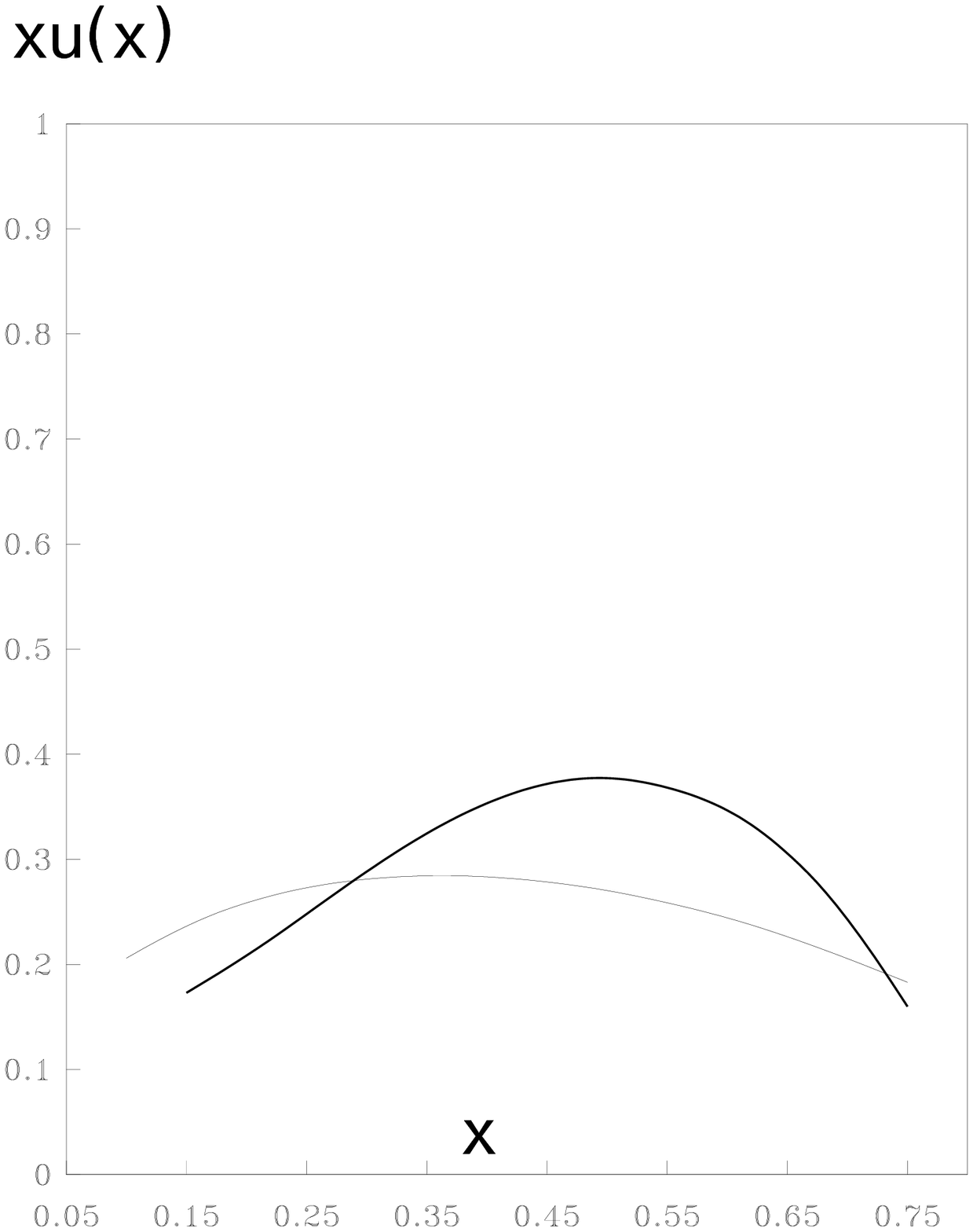}
\caption{ Quark distribution function in the pion (thick line) and 
the fit from [6] (thin line). }
\end{figure}

\begin{figure}
\epsfxsize=5cm
\epsfbox{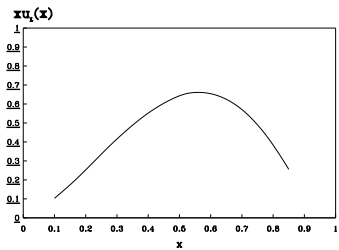}
\caption{ Quark distribution function for longitudinally polarizied 
$\rho$ meson }
\end{figure}

\begin{figure}
\epsfxsize=5cm
\epsfbox{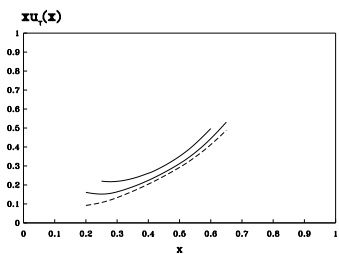}
\caption{Quark distribution function for transversally polarizied 
$\rho$-meson at three
choices of instanton radius $\rho_c=0.4, 0.5, 0.6$ fm.,
(curves from top to bottom, respectively)}
\end{figure}

\begin{figure}
\epsfxsize=5cm
\epsfbox{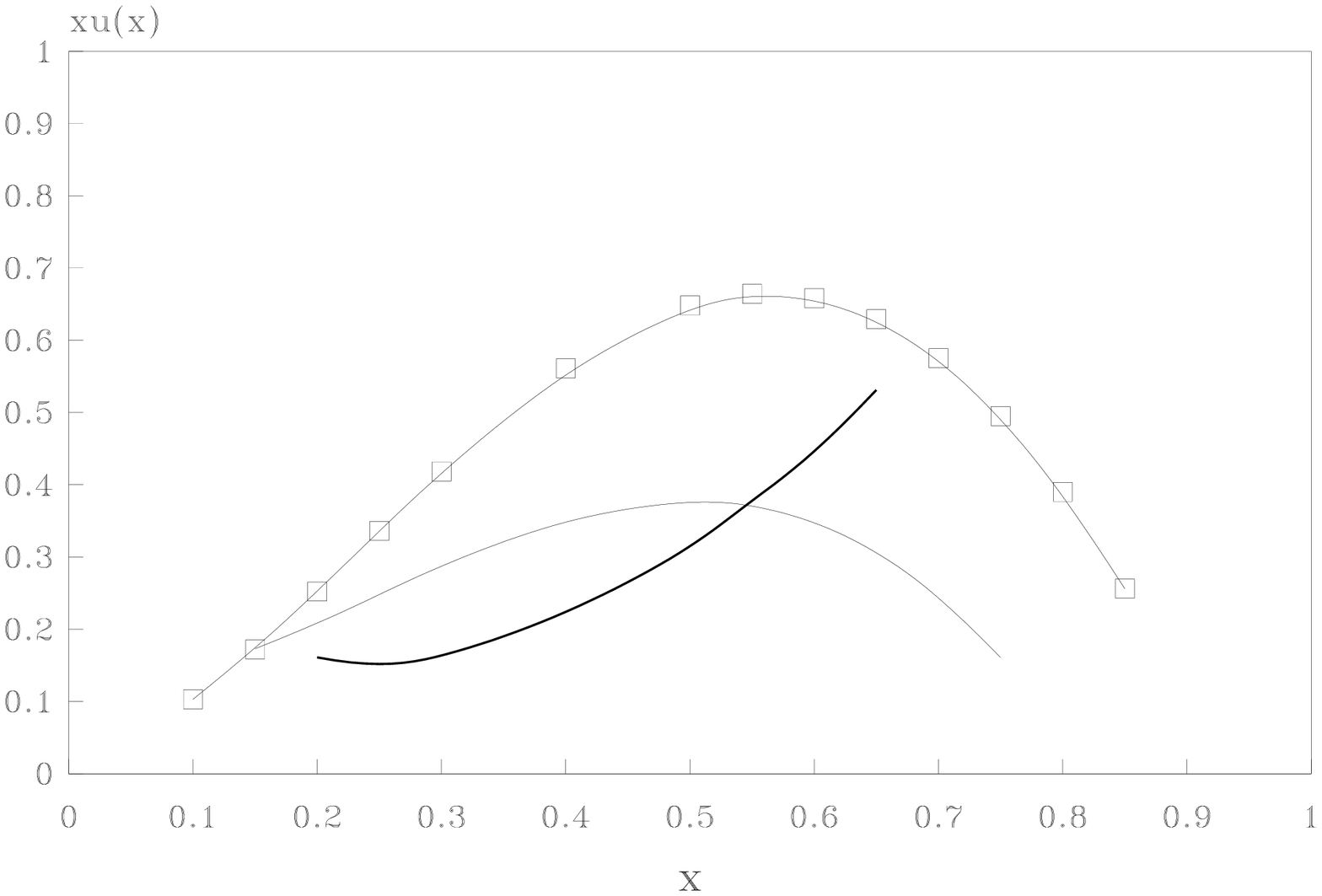}
\caption{Valence $u$ quark distributions for $\rho^T$-
(thick curve), $\rho^L$ (curve with squares) and $\pi$-meson (thin curve)}
\end{figure}
\newpage

\end{document}